\documentclass[12pt,reqno]{amsart}

\usepackage{amsmath,amsfonts,amsthm,amssymb}

\usepackage[a4paper,height=20cm,top=24mm,bottom=24mm,left=26mm,right=26mm]{geometry}

\usepackage[numbers,sort]{natbib}
\usepackage[breaklinks=true]{hyperref}

\usepackage[]{graphicx}
\usepackage{tabularx}     
\usepackage{booktabs}
\usepackage[table]{xcolor}
\numberwithin{equation}{section}

\newcommand{\ba}{\begin{eqnarray}}
\newcommand{\ea}{\end{eqnarray}}

\newcommand{\I}{\mathrm{i}}
\newcommand{\E}{\mathrm{e}}

\DeclareMathOperator{\Tr}{Tr}

\DeclareMathOperator{\ch}{ch}

\DeclareMathDelimiter{\Norm}{\mathord}{largesymbols}{"3E}{largesymbols}{"3E}

\DeclareMathOperator{\ord}{ord}
\allowdisplaybreaks[4]
\begin{document}
\baselineskip 16pt
\parskip 8pt
\sloppy

\begin{flushright}
YITP-10-74
\end{flushright}

\title[Twisted Elliptic Genus of $K3$]{Note on Twisted Elliptic Genus of $K3$ Surface}


\author[T. Eguchi]{Tohru \textsc{Eguchi}}

\author[K. Hikami]{Kazuhiro \textsc{Hikami}}


\address{Yukawa Institute for Theoretical Physics, Kyoto University,
  Kyoto 606--8502, Japan}
\email{
  \texttt{eguchi@yukawa.kyoto-u.ac.jp}
}

\address{Department of Mathematics, 
  Naruto University of Education,
  Tokushima 772-8502, Japan.}

\email{
  \texttt{KHikami@gmail.com}
}


\date{August 29, 2010.}

\begin{abstract}

We discuss the possibility of Mathieu group $M_{24}$ acting as symmetry group 
on the $K3$ elliptic genus as proposed recently by Ooguri, Tachikawa and
one of the present authors.
One way of testing this proposal is to derive the twisted elliptic genera for all  
conjugacy classes of $M_{24}$ so that we can 
determine  the unique decomposition of expansion coefficients of 
$K3$ elliptic genus into irreducible representations of $M_{24}$. 
In this paper 
we obtain all the hitherto unknown twisted elliptic genera and find a
strong evidence of Mathieu moonshine.

\end{abstract}


\keywords{
}

\subjclass[2000]{
}


\maketitle
\section{Introduction}

Let us consider the string theory compactified on the $K3$ surface.
It is well-known that string theory  on $K3$ has the symmetry of
$\mathcal{N}=4$ superconformal algebra.
In~\cite{EguOogTaoYan89a,EguchiHikami09a}
the elliptic genus
of $K3$  was expanded in terms of irreducible representations of $\mathcal{N}=4$
superconformal algebra 
\begin{align}
  Z_{K3}(z;\tau)
  &=8\left[\left({\theta_{10}(z;\tau)\over \theta_{10}(0;\tau)}\right)^2+
    \left({\theta_{00}(z;\tau)\over \theta_{00}(0;\tau)}\right)^2+
    \left({\theta_{01}(z;\tau)\over \theta_{01}(0;\tau)}\right)^2\right]\nonumber\\
  &=
  20\, \ch^{\widetilde{R}}_{h=\frac{1}{4},\ell=0}(z;\tau)
  -2\,\ch^{\widetilde{R}}_{h=\frac{1}{4},\ell=\frac{1}{2}}(z;\tau)
  +\sum_{n=1}^{\infty}
  A(n) \,
  \ch^{\widetilde{R}}_{h=n+\frac{1}{4},\ell=\frac{1}{2}}(z;\tau) .
  \label{Z_decompose}
\end{align}
In the RHS the first two terms denote characters of short (BPS)
representations of $\mathcal{N}=4$ algebra (with isospin $\ell=0$,
$1/2$),
and
an infinite series
denotes the sum over  
long (non-BPS) representations.
A new discovery was made in~\cite{EgucOoguTach10a}: expansion coefficients $A(n)$ agree with the
dimensions of irreducible or reducible dimensions of representations
of the Mathieu group $M_{24}$,
\begin{equation}
  \label{l1_coefficient_massive}
  \resizebox{\textwidth}{!}{$
  \begin{array}{c|cccccccccc}
     n & 1 & 2 & 3 & 4 & 5 & 6 & 7 & 8  &\cdots\\
    \hline
    A(n)& 2\times45 & 2\times 231 & 2\times 770 & 2\times 2277 & 2\times 5796 & 2\times 13915 &
     2\times 30843 & 2\times 65550 & \cdots
  \end{array}
$}
\end{equation}
In fact $M_{24}$ has 26 representations with dimensions
\begin{align*}
\{
&1,23,252,253,1771,3520,45,\overline{45},990,\overline{990},1035,\overline{1035},1035',231,\overline{231},
\\
&770,\overline{770},483,1265,2024,2277,3312,5313,5796,5544,10395\}.
\end{align*}
Here the pairs of
$45$,
$990$,
$1035$,
$231$,
$770$-dimensional representations are complex conjugate of each other. 
At the level $n=6$ we have a decomposition into irreducible
representations  $13915=3520+10395$ and at $n=7$ 
$30843=10395+5796+5544+5313+2024+1171$ etc.

Explicitly characters of short and long representations (in $\tilde{R}$ sector) are given by~\cite{EgucTaor88b}
\begin{gather}
  \ch^{\widetilde{R}}_{h={1\over      4},\ell=0}(z;\tau)={
    \left[ \theta_{11}(z;\tau) \right]^2\over
    \left[ \eta(\tau) \right]^3}\mu(z;\tau),
  \\[2mm]
  \ch^{\widetilde{R}}_{h=\frac{1}{4},\ell=\frac{1}{2}}(z;\tau)
  + 2 \,
  \ch^{\widetilde{R}}_{h=\frac{1}{4},\ell=0}(z;\tau)
  =
  q^{-\frac{1}{8}} \, \frac{
    \left[ \theta_{11}(z;\tau) \right]^2}{
    \left[ \eta(\tau) \right]^3
  } ,
  \\[2mm]
  \mu(z;\tau)=
  \frac{\I \, \E^{\pi \I z}}{
    \theta_{11}(z;\tau)
  }
  \sum_{n \in \mathbb{Z}}
  (-1)^n \,
  \frac{q^{\frac{1}{2} n (n+1)} \, \E^{2 \pi \I n z}}{
    1 - q^n \, \E^{2 \pi \I z}
  }  ,
\end{gather}
and
\begin{equation}
  \ch^{\widetilde{R}}_{h,\ell={1\over 2}}(z;\tau)
  =
  q^{h-{3\over 8}} \, {\left[ \theta_{11}(z;\tau) \right]^2\over
    \left[ \eta(\tau) \right]^3}.
\end{equation}
It is known that the above series $\mu(z;\tau)$ is a typical example of a Mock theta function~\cite{Zweg02Thesis}.
Expansion coefficients $A(n)$ are given in terms of $\mu$ at half-periods 
\begin{equation}
  -q^{{1\over 8}} \, \Sigma(\tau)\equiv -2+\sum_{n=1}^{\infty}
  A(n) \, q^n
  =
  8\hskip-5mm \sum_{w\in \{{1\over 2},{1+\tau\over 2},{\tau\over 2}\}} \,
  \hskip-5mm \mu(w;\tau).
  \label{Sigma}
\end{equation}
Prefactor $q^{1/8}$ above 
is for convenience. As studied in detail in~\cite{EguchiHikami09a}, $\Sigma(\tau)$ is a
mock theta function whose shadow~\cite{Zagier08a} is
$ \left[ \eta(\tau) \right]^3$.

Now consider a graded vector space 
\begin{equation*}
  \sum_{n=1}^{\infty}V(n)\,\, q^n
\end{equation*}
where the space $V(n)$ has a dimension $\dim V(n)=A(n)$. 
Since the dimension of the representation space is given by the trace of the identity element, 
we may rewrite the sum (\ref{Sigma}) as
\begin{equation}
  - q^{\frac{1}{8}} \,
  \Sigma(\tau)=-2+\sum_{n=1}^\infty\Tr_{V(n)}1\cdot  q^n.
\end{equation}

Twisted elliptic  genus is defined instead by considering an arbitrary
group element $g$
\begin{equation}
  -q^{\frac{1}{8}} \,\Sigma_g(\tau)
  =-2 +
  \sum_{n=1}^\infty A_g(n) \,  q^n ,
\end{equation}
where
\begin{equation}
  A_g(n) = \Tr_{V(n)} g .
  \label{define_A_g}
\end{equation}
Since the trace of $g$ depends only on its conjugacy class, 
there exists a twisted  elliptic genus
$Z_g(z;\tau)$ corresponding to each conjugacy class.
As a generalization of~\eqref{Z_decompose}, we define
the twisted elliptic genus by the decomposition
\begin{align}
  Z_g(z;\tau)
  & =
  \left( \chi_g - 4 \right) \,
  \ch^{\widetilde{R}}_{h=\frac{1}{4},\ell=0}(z;\tau)
  - 2 \,
  \ch^{\widetilde{R}}_{h=\frac{1}{4},\ell=\frac{1}{2}}(z;\tau)
  +
  \sum_{n=1}^\infty
  A_g(n) \,
  \ch^{\widetilde{R}}_{h=n+\frac{1}{4},\ell=\frac{1}{2}}(z;\tau) 
  \nonumber
  \\
  &=
  \chi_g \,
  \ch^{\widetilde{R}}_{h=\frac{1}{4},\ell=0}(z;\tau)
  -
  \frac{
    \left[ \theta_{11}(z;\tau) \right]^2}{
    \left[\eta(\tau) \right]^3} \,
  \Sigma_g(\tau)  ,
  \label{twisted_decompose}
\end{align}
where $\chi_g \in \mathbb{Z}$ is the Witten index $Z_g(z=0;\tau)$.
The $q$-series $\Sigma_g(\tau)$ is thus the analogue of the
McKay--Thompson series of
the monstrous moonshine~\cite{ConwayNorton79}.

The Mathieu group
$M_{24}$ has the following $26$ conjugacy classes: (we use the
ATLAS naming of the conjugacy classes~\cite{ATLAS85}.
See
Table~\ref{tab:class})
\begin{align}
  &\text{type I}:\,\,
  \mathrm{1A, 2A, 3A, 5A, 4B, 7A, 7B, 8A, 6A, 11A, 15A, 15B, 14A, 14B, 23A, 23B,}
  \label{typeI}\\
  &\text{type II}:\,\,
  \mathrm{12B, 6B, 4C, 3B, 2B, 10A, 21A, 21B, 4A, 12A.}
  \label{typeII}
\end{align}
Elements of the conjugacy classes of the type~I~\eqref{typeI}
fix 1 element out of 24 and these classes may be  considered as 
conjugacy classes of the subgroup $M_{23}$.
See the cycle
representation of these  classes in Table~\ref{tab:class}.
On the other hand, classes of the type~II~\eqref{typeII}
do not have a fixed element and are regarded as intrinsic elements of
$M_{24}$.
It turns out that
the twisted elliptic genera of these two types of conjugacy classes
have a qualitatively different behavior.
Especially the Witten index~$\chi_g$ in~\eqref{twisted_decompose} vanishes 
if and only if $g \in$ type~II;
\begin{equation}
  \begin{array}{c||rrrrrrrrrrrrr}
    g &
    \mathrm{1A} & \mathrm{2A} & \mathrm{3A} & \mathrm{5A} &
    \mathrm{4B} &  \mathrm{7A} & \mathrm{8A} & \mathrm{6A} &
    \mathrm{11A} & \mathrm{15A} & \mathrm{14A} & \mathrm{23A} &
    \text{others}
    \\
    \hline
    \chi_g &
    24 & 8 & 6 & 4 & 4 & 3 & 2 & 2 & 2 & 1 & 1 & 1 & 0
  \end{array}
\end{equation}
It is observed~\cite{MCheng10a,GabeHoheVolp10a} that $\chi_g$ is
related to the first and the second rows of the character table of $M_{24}$
in Table~\ref{tab:M24};
$\chi_{1A}=1+23$,
$\chi_{2A}=1+7$, and so on.
In the character decomposition~\eqref{twisted_decompose}, 
$A_g(n)$ is the Fourier coefficient of mock theta functions if
$g\in \text{type~I}$
and of modular form if $g \in \text{type~II}$.
Note  that twisted elliptic
genera for the pairs, 
$(\mathrm{7A, 7B})$,
$(\mathrm{15A, 15B})$,
$(\mathrm{14A, 14B})$,
$(\mathrm{23A, 23B})$,
$(\mathrm{21A, 21B})$,
are equal to each other. 

Twisted elliptic
genera of the conjugacy classes of type~I have already been obtained in the literature \cite{MCheng10a,GabeHoheVolp10a}. 
On the other hand, twisted elliptic
genera of type~II are yet largely
unknown.
In this paper we obtain all 
the twisted elliptic genera of type~II and then use the
character formula of the Mathieu group to derive the 
coefficients of the decomposition of  $K3$ elliptic genus into a sum
of irreducible representations of $M_{24}$.
We have checked that we always obtain the positive integral
coefficients in the decomposition up to $q^{600}$.
We thus provide a very strong support for the Mathieu moonshine conjecture.\footnote{Very recently a preprint~\cite{GabeHoheVolp10b} by M.Gaberdiel, S.Hohenegger and R. Volpato has appeared which has a substantial overlap with  
the present paper.}

\section{Twisted elliptic genus of the set type~I}
\label{sec:M23}

Twisted elliptic
genera of the type~I
has been obtained
previously~\cite{DavidJatkarSen06,MCheng10a,GabeHoheVolp10a}.
Type~I genera for basic classes,
$p\mathrm{A}$ ($p=2,3,5,7$) were discussed by A.Sen and his
collaborators~\cite{JatkarSen05,DavidJatkarSen06} 
(also~\cite{GovinKrish09a})
in connection with
the counting problem of ${1\over 4}$ BPS monopoles and dyons.

We first introduce the standard notation in the theory of Jacobi forms~\cite{EichZagi85}
\begin{equation}
\phi_{0,1}(z;\tau)=4\left[\left({\theta_{10}(z;\tau)\over \theta_{10}(0;\tau)}\right)^2+
\left({\theta_{00}(z;\tau)\over \theta_{00}(0;\tau)}\right)^2+
\left({\theta_{01}(z;\tau)\over \theta_{01}(0;\tau)}\right)^2\right],
\end{equation}
and
\begin{equation}
\phi_{-2,1}(z;\tau)=-{\left[ \theta_{11}(z;\tau) \right]^2\over 
  \left[\eta(\tau)\right]^6}.
\end{equation}
Here the Jacobi theta functions are defined in
Appendix~\ref{sec:theta}.
$\phi_{M,N}$ denotes a Jacobi form with weight $M$ and index
$N$.
We also use  the Eisenstein series
\begin{equation}
  \phi_2^{(N)}(\tau)
  =
  {24\over N-1}\,
  q \, \partial_q \log\left({\eta(N\tau)\over
      \eta(\tau)}\right)
  =1+{24\over N-1}\sum_{k=1}^\infty\sigma_1(k)(q^k
  -N \, q^{N k}).
\end{equation}

The elliptic genus for $K3$ is given by~\cite{KawaYamaYang94a,EguOogTaoYan89a}
\begin{equation}
 Z_{K3}(z;\tau)=Z_{\mathrm{1A}}(z;\tau) 
  = 2 \, \phi_{0,1}(z;\tau).
\end{equation}
Note that the class 
$\mathrm{1A}$ consists of the identity element and hence
$Z_{\mathrm{1A}}$ is the original untwisted elliptic genus. 

In the case of classes $p\mathrm{A}$
($p=2,3,5,7$) there is a general formula for the twisted elliptic
genera~\cite{MCheng10a}
\begin{equation}
  Z_{p\mathrm{A}}(z;\tau)
  =
  {2\over p+1} \,\phi_{0,1}(\tau)
  +{2p\over p+1} \,\phi_2^{(p)}(\tau) \, \phi_{-2,1}(z;\tau) .
\end{equation}
Explicitly we have
\begin{gather}
  \begin{aligned}[t]
    Z_{\mathrm{2A}}(z;\tau)
    & =
    \frac{2}{3} \, \phi_{0,1}(z;\tau)
    + \frac{4}{3} \, \phi_2^{(2)}(\tau) \,
    \phi_{-2,1}(z;\tau), 
  \end{aligned}
  \\
  \begin{aligned}[t]
    Z_{\mathrm{3A}}(z;\tau)
     & =
    \frac{1}{2} \, \phi_{0,1}(z;\tau)
    + \frac{3}{2} \, \phi_{2}^{(3)}(\tau) \, \phi_{-2,1}(z;\tau),
  \end{aligned}
  \\
  \begin{aligned}[t]
    Z_{\mathrm{5A}}(z;\tau)
   & = \frac{1}{3} \, \phi_{0,1}(z;\tau)
    +
    \frac{5}{3} \, \phi_2^{(5)}(\tau) \, \phi_{-2,1}(z;\tau),
  \end{aligned}
  \\
  \begin{aligned}[t]
    Z_{\mathrm{7A}}(z;\tau)
  & = \frac{1}{4} \, \phi_{0,1}(z;\tau)
    +
    \frac{7}{4} \, \phi_2^{(7)}(\tau) \, \phi_{-2,1}(z;\tau).
  \end{aligned}
\end{gather}
Twisted elliptic genera for other classes are given by~\cite{MCheng10a,GabeHoheVolp10a}
\begin{gather}
  \begin{aligned}[t]
    Z_{\mathrm{4B}}(z;\tau)
   & =
    \frac{1}{3} \, \phi_{0,1}(z;\tau)
    + \left(
      - \frac{1}{3} \, \phi_2^{(2)}(\tau)
      + 2 \, \phi_2^{(4)}(\tau)
    \right) \, \phi_{-2,1}(z;\tau),
  \end{aligned}
  \\
  \begin{aligned}[t]
    Z_{\mathrm{6A}}(z;\tau)
    & =
    \frac{1}{6} \, \phi_{0,1}(z;\tau)
    + \left(
      - \frac{1}{6} \, \phi_2^{(2)}(\tau)
      - \frac{1}{2} \, \phi_2^{(3)}(\tau)
      + \frac{5}{2} \, \phi_2^{(6)}(\tau)
    \right) \, \phi_{-2,1}(z;\tau),
  \end{aligned}
  \\
  \begin{aligned}[t]
    Z_{\mathrm{8A}}(z;\tau)
    & =
    \frac{1}{6} \, \phi_{0,1}(z;\tau)
    + \left(
      - \frac{1}{2} \, \phi_2^{(4)}(\tau)
      + \frac{7}{3} \, \phi_2^{(8)}(\tau)
    \right) \, \phi_{-2,1}(z;\tau),
  \end{aligned}
  \\
  \begin{aligned}[t]
    Z_{\mathrm{11A}}(z;\tau)
     & =
    \frac{1}{6} \, \phi_{0,1}(z;\tau)
    + \left(
      \frac{11}{6} \, \phi_2^{(11)}(\tau)
      - \frac{22}{5} \, 
      \left[ \eta(\tau) \, \eta(11 \, \tau) \right]^2
    \right) \, \phi_{-2,1}(z;\tau),
  \end{aligned}
  \\
  \begin{aligned}[t]
    Z_{\mathrm{14A}}(z;\tau)
    & =
    \frac{1}{12} \, \phi_{0,1}(z;\tau)
    + \left(
      -\frac{1}{36} \, \phi_2^{(2)}(\tau)
      -\frac{7}{12} \, \phi_2^{(7)}(\tau)
      +\frac{91}{36} \, \phi_2^{(14)}(\tau)
      \right.
      \\
      & \qquad \qquad \qquad \qquad \left.
      - \frac{14}{3} \, 
      \eta(\tau) \, \eta(2 \, \tau)  \, \eta(7 \,\tau)
      \, \eta(14 \, \tau)
    \right) \, \phi_{-2,1}(z;\tau),
  \end{aligned}
  \\
  \begin{aligned}[t]
    Z_{\mathrm{15A}}(z;\tau)
       & =
    \frac{1}{12} \, \phi_{0,1}(z;\tau)
    + \left(
      -\frac{1}{16} \, \phi_2^{(3)}(\tau)
      -\frac{5}{24} \, \phi_2^{(5)}(\tau)
      +\frac{35}{16} \, \phi_2^{(15)}(\tau)
      \right.
      \\
      & \qquad \qquad \qquad \qquad \left.
      - \frac{15}{4} \, 
      \eta(\tau) \, \eta(3 \, \tau)  \, \eta(5 \,\tau)
      \, \eta(15 \, \tau)
    \right) \, \phi_{-2,1}(z;\tau),
  \end{aligned}
  \\
  \begin{aligned}
    Z_{\mathrm{23A}}(z;\tau)
    & =
    \frac{1}{12} \, \phi_{0,1}(z;\tau)
    +
    \left(
      \frac{23}{12} \, \phi_2^{(23)}(\tau)
      -
      \frac{23}{22} \, f_{23,1}(\tau)
      - \frac{161}{22} \, f_{23,2}(\tau)
    \right) \,
    \phi_{-2,1}(z;\tau).
  \end{aligned}
\end{gather}
In the class 23A
we have used the 
newforms~\cite{DZagier92a}
\begin{align*}
  f_{23,1}(\tau)
  &=
  2 \, q - q^2 - q^4 - 2 \, q^5 - 5 \, q^6 + 2 \, q^7 
  + 4 \,  q^9 + 6 \, q^{10} - 6 \, q^{11} +   5 \, q^{12} + 6 \, q^{13}
  \\
  & \qquad \qquad 
  + 4 \, q^{14} - 10 \, q^{15} - 3  \, q^{16} 
  + 6 \, q^{17} - 2 \, q^{18} -   4 \, q^{19}+\cdots,
  \\[2mm]
  f_{23,2}(\tau)
  & =
  q^2 - 2 \, q^3 - q^4 + 2 \, q^5 + q^6 + 2 \, q^7 - 2 \, q^8 - 2 \, q^{10} - 
  2 \, q^{11} + q^{12}
  \\
  & \qquad \qquad
  + 2 \, q^{15} + 3 \, q^{16} - 2 \, q^{17} + 2 \, q^{18}
  +\cdots.
\end{align*}

\section{Twisted elliptic genus of the set type~II}

Our main task in this paper is to obtain all the twisted elliptic genera belonging  to type~II.
Here, unfortunately there is no definite guiding principle. We have to make 
an educated guess for the candidate
elliptic  genera which reproduce the correct coefficients of lower order 
$q$-expansions (see Table~\ref{tab:Fourier}) and have the correct
weight and level as modular forms. 
 
By trial and error we have obtained the following elliptic genera
which are written in
the form of $\eta$-product;
\begin{align}
  Z_{\mathrm{2B}}(z;\tau)
  &=2\,{\eta(\tau)^8\over \eta(2\tau)^4}\,\phi_{-2,1}(z;\tau),
  \\[2mm]
  Z_{\mathrm{4A}}(z;\tau)
  &=2\,{\eta(2\tau)^8\over \eta(4\tau)^4}\,\phi_{-2,1}(z;\tau),
  \\[2mm]
  Z_{\mathrm{4C}}(z;\tau)
  &=2\,{\eta(\tau)^4 \, \eta(2\tau)^2\over
    \eta(4\tau)^2}\,\phi_{-2,1}(z;\tau),
  \\[2mm]
  Z_{\mathrm{3B}}(z;\tau)
  &=2\,{\eta(\tau)^6\over \eta(3\tau)^2}\,\phi_{-2,1}(z;\tau),
  \\[2mm]
  Z_{\mathrm{6B}}(z;\tau)
  &=2\,{\eta(\tau)^2 \, \eta(2\tau)^2 \, \eta(3\tau)^2\over
    \eta(6\tau)^2}\,\phi_{-2,1}(z;\tau),
  \\[2mm]
  Z_{\mathrm{12B}}(z;\tau)
  &=2\,{\eta(\tau)^4 \, \eta(4\tau) \, \eta(6\tau)\over
    \eta(2\tau) \, \eta(12\tau)}\,\phi_{-2,1}(z;\tau),
  \\[2mm]
  Z_{\mathrm{10A}}(z;\tau)
  &=2\,{\eta(\tau)^3 \, \eta(2\tau) \, \eta(5\tau)\over
    \eta(10\tau)}\,\phi_{-2,1}(z;\tau),
  \\[2mm]
  Z_{\mathrm{12A}}(z;\tau)
  &=2\,{\eta(\tau)^3 \, \eta(4\tau)^2 \,\eta(6\tau)^3\over
    \eta(2\tau) \, \eta(3\tau) \,\eta(12\tau)^2}\,\phi_{-2,1}(z;\tau).
\end{align}
For the class $\mathrm{21A}$ one has a linear combination of $\eta$-products
\begin{equation}
  Z_{\mathrm{21A}}(z;\tau)
  =
  \left(
    \frac{7}{3} \,
    \frac{
      \eta(\tau)^3 \,
      \eta(7 \, \tau)^3
    }{
      \eta(3 \, \tau) \, \eta(21 \, \tau)
    }
    - \frac{1}{3} \,
    \frac{ \eta(\tau)^6}{
      \eta(3 \, \tau)^2}   \right) \,
  \phi_{-2,1}(z;\tau).
\end{equation}
One sees
that $\Sigma_g(\tau)$ is the $\eta$-product which 
is modular on congruence subgroup $\Gamma_0(\ord(g))$ with character.

We note the following relation among the genera of  type~I and type~II
\begin{gather}
  Z_{\mathrm{2A}}(z;\tau) + Z_{\mathrm{2B}}(z;\tau)
  =2 \, Z_{\mathrm{4B}} (z;\tau),
  \\[2mm]
  Z_{\mathrm{4A}}(z;\tau) + Z_{\mathrm{4B}}(z;\tau)
  = 2 \, Z_{\mathrm{8A}}(z;\tau).
\end{gather}

\section{Mathieu moonshine}

In Table~\ref{tab:M24} the character formula for the Mathieu group
$M_{24}$ is presented.
We denote its elements as $\chi_R^{\,\,\,g}$ where $R$ runs over
irreducible 
representations and $g$ runs over 
conjugacy classes.
It is well-known that the character formula obeys the orthogonality
relation
\begin{equation}
  \sum_g
  n_g\,\chi_{R'}^{\,\,\,g}\,\overline{\chi}_{R}^{\,\,\,g}=
  |G|\,\delta_{RR'}
  \label{orthogonal}
\end{equation}
where $n_g$ is the number of elements in the conjugacy class $g$ and
$|G|$ is the order of the group $G$. 
Let us denote the multiplicity of
the representation $R$ in the decomposition of the $K3$ elliptic genus
at level $n$ as $c_R(n)$. We then obtain the value of the twisted
genus of the class $g$ 
at level $n$ as
\begin{equation}
  \sum_R c_R(n) \,
  \chi_R^{\,\,\,g}=A_g(n) ,
\label{value-twisted-genus}
\end{equation}
where $A_g(n)$ is defined in~\eqref{define_A_g}. Note that 
by choosing $g=\mathrm{1A}$
in~\eqref{value-twisted-genus}  we find
\begin{equation}
  \sum_R c_n(R) \, \chi_R^{\,\,\,\mathrm{1A}}=\sum_R c_n(R)
  \dim R=A(n)
  \end{equation}
In fact $c_R(n)$ is the multiplicity of representation $R$ at level
$n$.

If one uses the orthogonality relation~\eqref{orthogonal}, we can
invert the relation~\eqref{value-twisted-genus} and find a formula for
the multiplicities
\begin{equation}
  \sum_g {1\over |G|} \,
  n_g\,\overline{\chi}_R^{\,\,\,g} \,A_g(n)=c_R(n).
\end{equation}

We have checked by computer that the multiplicities $c_R(n)$ are
positive integers  for all representations up to $n=600$.
See Table~\ref{tab:multiplicity}.
This provides a very strong
support of the Mathieu moonshine conjecture.

\section{Entropy}
In Table~\ref{tab:Fourier}, we have tabulated the values of 
$A_g(n)$, the expansion coefficients of twisted genera 
$Z_g(z;\tau)$.
In the untwisted case ($g=\mathrm{1A}$)
we have applied the 
the method of Bringmann--Ono~\cite{BrinKOno06a}
and obatined the Poincar{\'e} series ~\cite{EguchiHikami09a}
\begin{equation}
  \label{K3_An}
  A(n) =
  \frac{- 2 \, \pi \, \I}{
    \left( 8 \, n - 1\right)^{\frac{1}{4}}
  }
  \sum_{c=1}^\infty
  \frac{1}{\sqrt{c}} \,
  I_{\frac{1}{2}}
  \left(
    \frac{\pi \, \sqrt{ 8 \, n -1}}{2 \, c}
  \right) \,
  \sum_{\substack{
      k \mod 4 c
      \\
      k^2= - 8 n +1 \mod 8c
    }}
  \left(
    \frac{-4}{k}
  \right) \,
  \E^{\frac{k}{2 c} \pi \I} ,
\end{equation}
where
$\left(\frac{-4}{\bullet}\right)$ is the Legendre symbol,
and $I$ denotes the Bessel function,
\begin{equation*}
  I_{\frac{1}{2}}(x)
  =
  \sqrt{\frac{2}{\pi \, x}} \,
  \sinh(x) .
\end{equation*}
We have identified the expotential growth of $\{ A(n) \}$ at large $n$ 
as the entropy of $K3$ surface
\begin{equation}
  S_{K3} = \log A(n)
  \sim
  2 \, \pi \, \sqrt{ \frac{1}{2} \, \left( n - \frac{1}{8} \right)} .
\end{equation}
See~\cite{EguchiHikami09b,EguchiHikami10a} for the discussion of entropy of
higher-dimensional complex manifolds with reduced holonomy.

In case of $g \in \text{type~I}$,
twisted elliptic genus $Z_g(z;\tau)$ 
is modular on the congruence subgroup
$\Gamma_0(\ord(g))$ of $SL(2; \mathbb{Z})$.
Correspondingly
the $q$-series $\Sigma_g(\tau)$ is a mock theta 
function on $\Gamma_0(\ord(g))$, and by using the same method as above 
the Fourier coefficients $A_g(n)$ are given by
\begin{equation}
  A_g(n) =
  \frac{- 2 \, \pi \, \I}{
    \left( 8 \, n - 1\right)^{\frac{1}{4}}
  }
  \sum_{\substack{
      c=1\\
    \ord(g)|c
  }}^\infty
  \frac{1}{\sqrt{c}} \,
  I_{\frac{1}{2}}
  \left(
    \frac{\pi \, \sqrt{ 8 \, n -1}}{2 \, c}
  \right) \,
  \sum_{\substack{
      k \mod 4 c
      \\
      k^2= - 8 n +1 \mod 8c
    }}
  \left(
    \frac{-4}{k}
  \right) \,
  \E^{\frac{k}{2 c} \pi \I} .
\end{equation}
See~\cite{EguchiHikami09a} where a case of $\Gamma_0(2)$ was studied.
The above formula shows that
the entropy $S_g$ of ``twisted'' $K3$ is given by
\begin{equation}
  S_g = \log \left| A_g(n) \right|
  \sim \frac{S_{K3}}{\ord(g)} .
\end{equation}
Thus the entropy of  twisted $K3$ is reduced by a factor
$1/\ord(g)$.
This coincides with the result of~\cite{ASen10b} that the
entropy of the
$\mathbb{Z}_N$ twisted CHL model is  $1/N$ times the entropy of the
untwisted model.

 \section{Discussions}
 
We have completed the analysis initiated in
\cite{MCheng10a,GabeHoheVolp10a} on Mathieu moonshine phenomenon by
providing all the twisted elliptic genera for $K3$ surface. 
Making use of them we are able to decompose uniquely the expansion
coefficients $\{A_g(n)\}$ into a sum of irreducible representations of
$M_{24}$.
We find that
the multiplicities of all irreducible representations are positive integers 
up to the level $n=600$.

 For the sake of illustration we present the decomposition at the level $n=98$;
 \begin{equation*}
   \begin{aligned}[b]
&1754939889054075390 \\
&= 7168167560\times 1+164868700882\times 23 +1806385660318\times 252 \\
&+1813554671156\times 253+12694876811718\times 1771+25232056046588\times 3520\\
&
+ 322568604932\times \left( 45+  \overline{45} \right)
+7096515632052\times \left( 990 + \overline{990} \right)
\\
&+7419084236984\times \left( 1035+ \overline{1035} \right)
 +7419083183322\times 1035'
\\
&+1655854140602\times \left( 231+ \overline{231} \right)
+5519511336942 \times \left( 770+ \overline{770} \right)
\\
&
 +3462239800920\times 483
 +9067771260936\times 1265
+ 14508430647818\times 2024
\\
&+16321986797048\times 2277
+ 23741069980370\times 3312
 +38084633405380\times 5313
\\
&+41546870254732\times 5796
 +39740484586192\times 5544 +74513414138524\times 10395
\end{aligned}
\end{equation*}
Thus the observation of \cite{EgucOoguTach10a} may well be proved to be true.  

We are, however, still very far from satisfactory understanding of the
origin of the symmetry of the Mathieu group~$M_{24}$.
As is well-known, there are special classes of $K3$ surfaces
which possess automorphism under subgroups
of~$M_{23}$~\cite{Mukai88a,SKondo98a}
(see~\cite{TaormWendl10a} for a recent result). 
Thus it appears that~$M_{24}$ emerges as an enhanced symmetry in
string theory. Hopefully the twisted genera we have obtained   offer
some clue in our search for the action of $M_{24}$ on the string
Hilbert space   in $K3$ compactification.

\section*{Acknowledgments}
We thank H.Ooguri and Y.Tachikawa for their collaboration at the early
stage of this work. 
K.H.  thanks D.Zagier for useful discussions at MFO, and participants
in ``Prospects in $q$-series and modular forms'' at Dublin for discussions.
Some computations are performed by use of SAGE~\cite{SAGE}.
Research of T.E. and K.H. are supported by Japan Ministry of
Education, Culture, Sports, Science and Technology under grant
No.~19GS0219, 
21654053, and 
22540069.

\appendix

\section{Jacobi Theta Functions}
\label{sec:theta}
The Jacobi theta functions are defined by
\begin{equation}
  \begin{aligned}
    \theta_{11}(z;\tau)
    & =
    \sum_{n \in \mathbb{Z}}
    q^{\frac{1}{2} \left( n+ \frac{1}{2} \right)^2} \,
    \E^{2 \pi \I \left(n+\frac{1}{2} \right) \,
      \left( z+\frac{1}{2} \right)
    }
    ,
    \\[2mm]
    \theta_{10}(z;\tau)
    & =
    \sum_{n \in \mathbb{Z}}
    q^{\frac{1}{2} \left( n + \frac{1}{2} \right)^2} \,
    \E^{2 \pi \I \left( n+\frac{1}{2} \right) z}
    ,
    \\[2mm]
    \theta_{00} (z;\tau)
    & =
    \sum_{n \in \mathbb{Z}}
    q^{\frac{1}{2} n^2} \,
    \E^{2 \pi \I  n  z}
    ,
    \\[2mm]
    \theta_{01} (z;\tau)
    & =
    \sum_{n \in \mathbb{Z}}
    q^{\frac{1}{2} n^2} \,
    \E^{2 \pi \I n \left( z+\frac{1}{2} \right) }
    .
  \end{aligned}
\end{equation}
Throughout this paper, we set $q=\E^{2 \pi \I \tau}$ with
$\tau$ in the upper half plane,
$\tau \in \mathbb{H}$.

\newpage 
\clearpage 
\begin{table}
  \newcolumntype{L}{>{$}l<{$}}
  \newcolumntype{C}{>{$}c<{$}}
  \rowcolors{2}{gray!22}{}
  \centering
  \resizebox{\textwidth}{!}{
  \begin{tabular}{c||C|L}
    \toprule
    conjugacy class & \text{cycle shape} &
    \\
    \midrule \midrule
    1A & 1^{24} & ()
    \\
    2A &1^8 \cdot 2^8 & 
    (1,8)(2,12)(4,15)(5,7)(9,22)(11,18)(14,19)(23,24)
    \\
    3A & 1^6 \cdot 3^6 &
    (3,18,20)(4,22,24)(5,19,17)(6,11,8)(7,15,10)(9,12,14)
    \\
    5A & 1^4 \cdot 5^4&
    (2,21,13,16,23)(3,5,15,22,14)(4,12,20,17,7)(9,18,19,10,24)
    \\
    4B &1^4 \cdot 2^2 \cdot 4^4 &
    (1,17,21,9)(2,13,24,15)(3,23)(4,14,5,8)(6,16)(12,18,20,22)
    \\
    7A &1^3 \cdot 7^3 &
    (1,17,5,21,24,10,6)(2,12,13,9,4,23,20)(3,8,22,7,18,14,19)
    \\
    7B&1^3 \cdot 7^3 &
    (1,21,6,5,10,17,24)(2,9,20,13,23,12,4)(3,7,19,22,14,8,18)
    \\
    8A &1^2 \cdot 2^1 \cdot 4^1 \cdot 8^2 &
    (1,13,17,24,21,15,9,2)(3,16,23,6)(4,22,14,12,5,18,8,20)(7,11)
    \\ 
    6A &1^2 \cdot 2^2 \cdot 3^2 \cdot 6^2 &
    (1,8)(2,24,11,12,23,18)(3,20,10)(4,15)(5,19,9,7,14,22)(6,16,13)
    \\
    11A&1^2 \cdot 11^2 &
    (1,3,10,4,14,15,5,24,13,17,18)(2,21,23,9,20,19,6,12,16,11,22)
    \\
    15A &1^1 \cdot 3^1 \cdot 5^1 \cdot 15^1 &
    (2,13,23,21,16)(3,7,9,5,4,18,15,12,19,22,20,10,14,17,24)(6,8,11)
    \\
    15B& 1^1 \cdot 3^1 \cdot 5^1 \cdot 15^1 &
    (2,23,16,13,21)(3,12,24,15,17,18,14,4,10,5,20,9,22,7,19)(6,8,11)
    \\
    14A &1^1 \cdot 2^1 \cdot 7^1 \cdot 14^1 &
    (1,12,17,13,5,9,21,4,24,23,10,20,6,2)(3,18,8,14,22,19,7)(11,15)
    \\
    14B& 1^1 \cdot 2^1 \cdot 7^1 \cdot 14^1 &
    (1,13,21,23,6,12,5,4,10,2,17,9,24,20)(3,14,7,8,19,18,22)(11,15)
    \\
    23A &1^1 \cdot 23^1 &
    (1,7,6,24,14,4,16,12,20,9,11,5,15,10,19,18,23,17,3,2,8,22,21)
    \\
    23B& 1^1 \cdot 23^1 &
    (1,4,11,18,8,6,12,15,17,21,14,9,19,2,7,16,5,23,22,24,20,10,3)
    \\
    \midrule
    12B &12^2 &
    (1,12,24,23,10,8,18,6,3,21,2,7)(4,9,11,15,13,16,20,5,22,17,14,19)
    \\
    6B &6^4 &
    (1,24,10,18,3,2)(4,11,13,20,22,14)(5,17,19,9,15,16)(6,21,7,12,23,8)
    \\
    4C &4^6 &
    (1,23,18,21)(2,12,10,6)(3,7,24,8)(4,15,20,17)(5,14,9,13)(11,16,22,19)
    \\
    3B &3^8 &
    (1,10,3)(2,24,18)(4,13,22)(5,19,15)(6,7,23)(8,21,12)(9,16,17)(11,20,14)
    \\
    2B &2^{12} &
    (1,8)(2,10)(3,20)(4,22)(5,17)(6,11)(7,15)(9,13)(12,14)(16,18)(19,23)(21,24)
    \\
    10A &2^2 \cdot 10^2 &
    (1,8)(2,18,21,19,13,10,16,24,23,9)(3,4,5,12,15,20,22,17,14,7)(6,11)
    \\
    21A &3^1 \cdot 21^1 &
    (1,3,9,15,5,12,2,13,20,23,17,4,14,10,21,22,19,6,7,11,16)(8,18,24)
    \\
    21B& 3^1 \cdot 21^1 &
    (1,12,17,22,16,5,23,21,11,15,20,10,7,9,13,14,6,3,2,4,19)(8,24,18)
    \\
    4A &2^4 \cdot 4^4 &
    (1,4,8,15)(2,9,12,22)(3,6)(5,24,7,23)(10,13)(11,14,18,19)(16,20)(17,21)
    \\
    12A &2^1 \cdot 4^1 \cdot 6^1 \cdot 12^1 &
    (1,15,8,4)(2,19,24,9,11,7,12,14,23,22,18,5)(3,13,20,6,10,16)(17,21)
    \\
    \bottomrule
  \end{tabular}
}
  \caption{Representatives of conjugacy classes.}
  \label{tab:class}
\end{table}
\clearpage 

\begin{table}
  \newcolumntype{R}{>{$}r<{$}}
  \rowcolors{2}{gray!22}{}
  \rotatebox[]{90}{
    \resizebox{0.95\textheight}{!}{
      \centering
        \begin{tabular}{RRRRRRRRRRRRRRRRRRRRRRRRRR}
          \toprule
          \text{1A} & \text{2A} & \text{3A} &\text{5A} & \text{4B}
          &\text{7A}&\text{7B}& \text{8A} & \text{6A} &
          \text{11A}&\text{15A} & \text{15B} & \text{14A} &
          \text{14B} & \text{23A} &\text{23B} & \text{12B}&
          \text{6B} & \text{4C} & \text{3B} & \text{2B} &\text{10A}
          & \text{21A} & \text{21B} & \text{4A} &\text{12A} \\
          \midrule \midrule
          1 & 1 & 1 & 1 & 1 & 1 & 1 & 1 & 1 & 1 & 1 & 1 & 1 & 1 & 1 & 1 & 1 & 1 & 1 & 1 & 1 & 1 & 1 & 1 & 1 & 1 \\
          23 & 7 & 5 & 3 & 3 & 2 & 2 & 1 & 1 & 1 & 0 & 0 & 0 & 0 & 0 & 0 & -1 & -1 & -1 & -1 & -1 & -1 & -1 & -1 & -1 &
          -1 \\
          252 & 28 & 9 & 2 & 4 & 0 & 0 & 0 & 1 & -1 & -1 & -1 & 0 & 0 & -1 & -1 & 0 & 0 & 0 & 0 & 12 & 2 & 0 & 0 & 4 & 1
          \\
          253 & 13 & 10 & 3 & 1 & 1 & 1 & -1 & -2 & 0 & 0 & 0 & -1 & -1 & 0 & 0 & 1 & 1 & 1 & 1 & -11 & -1 & 1 & 1 & -3 &
          0 \\
          1771 & -21 & 16 & 1 & -5 & 0 & 0 & -1 & 0 & 0 & 1 & 1 & 0 & 0 & 0 & 0 & -1 & -1 & -1 & 7 & 11 & 1 & 0 & 0 & 3 &
          0 \\
          3520 & 64 & 10 & 0 & 0 & -1 & -1 & 0 & -2 & 0 & 0 & 0 & 1 & 1 & 1 & 1 & 0 & 0 & 0 & -8 & 0 & 0 & -1 & -1 & 0 &
          0 \\
          45 & -3 & 0 & 0 & 1 & e_7^+ & e_7^- & -1 & 0 & 1 & 0 & 0 & -e_7^+ & -e_7^- & -1 & -1 & 1 &
          -1 & 1 & 3 & 5 & 0 & e_7^- & e_7^+ & -3 & 0 \\
          \overline{45} & -3 & 0 & 0 & 1 & e_7^- & e_7^+ & -1 & 0 & 1 & 0 & 0 & -e_7^- & -e_7^+ & -1 & -1 & 1 &
          -1 & 1 & 3 & 5 & 0 & e_7^+ & e_7^- & -3 & 0 \\
          990 & -18 & 0 & 0 & 2 & e_7^+ & e_7^- & 0 & 0 & 0 & 0 & 0 & e_7^+ & e_7^- & 1 & 1 & 1 & -1
          & -2 & 3 & -10 & 0 & e_7^- & e_7^+ & 6 & 0 \\
          \overline{990} & -18 & 0 & 0 & 2 & e_7^- & e_7^+ & 0 & 0 & 0 & 0 & 0 & e_7^- & e_7^+ & 1 & 1 & 1 & -1
          & -2 & 3 & -10 & 0 & e_7^+ & e_7^- & 6 & 0 \\
          1035 & -21 & 0 & 0 & 3 & 2 e_7^+ & 2 e_7^- & -1 & 0 & 1 & 0 & 0 & 0 & 0 & 0 & 0 & -1 & 1 & -1 & -3 &
          -5 & 0 & -e_7^- & -e_7^+ & 3 & 0 \\
          \overline{1035} & -21 & 0 & 0 & 3 & 2 e_7^- & 2 e_7^+ & -1 & 0 & 1 & 0 & 0 & 0 & 0 & 0 & 0 & -1 & 1 & -1 & -3 &
          -5 & 0 & -e_7^+ & -e_7^- & 3 & 0 \\
          1035^\prime & 27 & 0 & 0 & -1 & -1 & -1 & 1 & 0 & 1 & 0 & 0 & -1 & -1 & 0 & 0 & 0 & 2 & 3 & 6 & 35 & 0 & -1 & -1 & 3 &
          0 \\
          231 & 7 & -3 & 1 & -1 & 0 & 0 & -1 & 1 & 0 & e_{15}^+ & e_{15}^- & 0 & 0 & 1 & 1 & 0 & 0 & 3 & 0 & -9 & 1 &
          0 & 0 & -1 & -1 \\
          \overline{231} & 7 & -3 & 1 & -1 & 0 & 0 & -1 & 1 & 0 & e_{15}^- & e_{15}^+ & 0 & 0 & 1 & 1 & 0 & 0 & 3 & 0 & -9 & 1 &
          0 & 0 & -1 & -1 \\
          770 & -14 & 5 & 0 & -2 & 0 & 0 & 0 & 1 & 0 & 0 & 0 & 0 & 0 & e_{23}^+ & e_{23}^- & 1 & 1 & -2 & -7 & 10 & 0
          & 0 & 0 & 2 & -1 \\
          \overline{770} & -14 & 5 & 0 & -2 & 0 & 0 & 0 & 1 & 0 & 0 & 0 & 0 & 0 & e_{23}^- & e_{23}^+ & 1 & 1 & -2 & -7 & 10 & 0
          & 0 & 0 & 2 & -1 \\
          483 & 35 & 6 & -2 & 3 & 0 & 0 & -1 & 2 & -1 & 1 & 1 & 0 & 0 & 0 & 0 & 0 & 0 & 3 & 0 & 3 & -2 & 0 & 0 & 3 & 0 \\
          1265 & 49 & 5 & 0 & 1 & -2 & -2 & 1 & 1 & 0 & 0 & 0 & 0 & 0 & 0 & 0 & 0 & 0 & -3 & 8 & -15 & 0 & 1 & 1 & -7 &
          -1 \\
          2024 & 8 & -1 & -1 & 0 & 1 & 1 & 0 & -1 & 0 & -1 & -1 & 1 & 1 & 0 & 0 & 0 & 0 & 0 & 8 & 24 & -1 & 1 & 1 & 8 &
          -1 \\
          2277 & 21 & 0 & -3 & 1 & 2 & 2 & -1 & 0 & 0 & 0 & 0 & 0 & 0 & 0 & 0 & 0 & 2 & -3 & 6 & -19 & 1 & -1 & -1 & -3 &
          0 \\
          3312 & 48 & 0 & -3 & 0 & 1 & 1 & 0 & 0 & 1 & 0 & 0 & -1 & -1 & 0 & 0 & 0 & -2 & 0 & -6 & 16 & 1 & 1 & 1 & 0 & 0
          \\
          5313 & 49 & -15 & 3 & -3 & 0 & 0 & -1 & 1 & 0 & 0 & 0 & 0 & 0 & 0 & 0 & 0 & 0 & -3 & 0 & 9 & -1 & 0 & 0 & 1 & 1
          \\
          5796 & -28 & -9 & 1 & 4 & 0 & 0 & 0 & -1 & -1 & 1 & 1 & 0 & 0 & 0 & 0 & 0 & 0 & 0 & 0 & 36 & 1 & 0 & 0 & -4 &
          -1 \\
          5544 & -56 & 9 & -1 & 0 & 0 & 0 & 0 & 1 & 0 & -1 & -1 & 0 & 0 & 1 & 1 & 0 & 0 & 0 & 0 & 24 & -1 & 0 & 0 & -8 &
          1 \\
          10395 & -21 & 0 & 0 & -1 & 0 & 0 & 1 & 0 & 0 & 0 & 0 & 0 & 0 & -1 & -1 & 0 & 0 & 3 & 0 & -45 & 0 & 0 & 0 & 3 &
          0
          \\
          \bottomrule
        \end{tabular}
        }}
  \caption{Character table of the Mathieu group $M_{24}$.
  Here we have used
  $e_p^\pm = \frac{1}{2} \,
  \left( \pm \sqrt{-p} -1 \right)$.
}
  \label{tab:M24}
\end{table}

\begin{table}
  \newcolumntype{R}{>{$}r<{$}}
  \rowcolors{2}{gray!22}{}
  \rotatebox[]{90}{
    \resizebox{0.95\textheight}{!}{
    \centering
    \begin{tabular}{R||RRRRRRRRRRRR|RRRRRRRRR}
      \toprule
     n & \text{1A} & \text{2A} & \text{3A} & \text{5A}
      & \text{4B} & \text{7A} & \text{8A} & \text{6A}
      & \text{11A} & \text{15A} & \text{14A} & \text{23A}
        &\text{12B} & \text{6B} & \text{4C} & \text{3B}
        & \text{2B}    & \text{10A} & \text{21A} & \text{4A}
        & \text{12A}
        \\
        \midrule \midrule
1&90&-6&0&0&2&-1&-2&0&2&0&1&-2&2&-2&2&6&10&0&-1&-6&0\\
2&462&14&-6&2&-2&0&-2&2&0&-1&0&2&0&0&6&0&-18&2&0&-2&-2\\
3&1540&-28&10&0&-4&0&0&2&0&0&0&-1&2&2&-4&-14&20&0&0&4&-2\\
4&4554&42&0&-6&2&4&-2&0&0&0&0&0&0&4&-6&12&-38&2&-2&-6&0\\
5&11592&-56&-18&2&8&0&0&-2&-2&2&0&0&0&0&0&0&72&2&0&-8&-2\\
6&27830&86&20&0&-2&-2&2&-4&0&0&2&0&0&0&6&-16&-90&0&-2&6&0\\
7&61686&-138&0&6&-10&2&-2&0&-2&0&2&0&-2&-2&-2&30&118&-2&2&6&0\\
8&131100&188&-30&0&4&-3&0&2&2&0&-1&0&0&0&-12&0&-180&0&0&-4&2\\
9&265650&-238&42&-10&10&0&-2&2&0&2&0&0&-2&6&10&-42&258&-2&0&-14&-2\\
10&521136&336&0&6&-8&0&-4&0&0&0&0&2&-2&2&16&42&-352&-2&0&0&0\\
11&988770&-478&-60&0&-14&6&2&-4&2&0&-2&0&0&0&-6&0&450&0&0&18&0\\
12&1830248&616&62&8&8&0&0&-2&2&2&0&0&2&-6&-16&-70&-600&0&0&-8&-2\\
13&3303630&-786&0&0&22&-6&2&0&0&0&-2&2&0&-4&6&84&830&0&0&-18&0\\
14&5844762&1050&-90&-18&-6&0&2&6&0&0&0&2&0&0&18&0&-1062&-2&0&10&-2\\
15&10139734&-1386&118&4&-26&-4&-2&6&0&-2&0&0&2&2&-10&-110&1334&4&2&22&-2\\
16&17301060&1764&0&0&12&0&0&0&-4&0&0&0&2&6&-28&126&-1740&0&0&-12&0\\
17&29051484&-2212&-156&14&28&0&-4&-4&0&-1&0&0&0&0&12&0&2268&-2&0&-36&0\\
18&48106430&2814&170&0&-18&8&-2&-6&-2&0&0&-2&2&-6&38&-166&-2850&0&2&14&2\\
19&78599556&-3612&0&-24&-36&0&0&0&2&0&0&0&-2&-6&-20&210&3540&0&0&36&0\\
20&126894174&4510&-228&14&14&-6&-2&4&0&2&2&0&0&0&-42&0&-4482&-2&0&-18&0\\
21&202537080&-5544&270&0&48&4&4&6&-2&0&0&0&-2&6&16&-282&5640&0&-2&-40&2\\
22&319927608&6936&0&18&-16&-7&4&0&0&0&-1&0&0&4&48&300&-6968&2&-1&24&0\\
23&500376870&-8666&-360&0&-58&0&-2&-8&4&0&0&2&0&0&-18&0&8550&0&0&54&0\\
24&775492564&10612&400&-36&28&0&0&-8&0&0&0&0&0&-8&-60&-392&-10556&4&0&-28&-4\\
25&1191453912&-12936&0&12&64&12&-4&0&0&0&0&0&2&-10&32&462&13064&4&0&-72&0\\
26&1815754710&15862&-510&0&-34&0&-6&10&0&0&0&-1&0&0&78&0&-15930&0&0&22&-2\\
27&2745870180&-19420&600&30&-76&-10&4&8&-2&0&-2&0&0&8&-36&-600&19268&-2&2&84&0\\
28&4122417420&23532&0&0&36&2&0&0&0&0&-2&0&0&12&-84&660&-23460&0&2&-36&0\\
29&6146311620&-28348&-762&-50&100&-6&4&-10&-2&-2&2&0&0&0&36&0&28548&-2&0&-92&-2\\
30&9104078592&34272&828&22&-40&0&4&-12&4&-2&0&0&0&-8&96&-840&-34352&-2&0&48&0\\
31&13401053820&-41412&0&0&-116&0&-4&0&0&0&0&-2&-2&-10&-44&966&41180&0&0&108&0\\
32&19609321554&49618&-1062&34&50&18&2&10&-2&-2&2&0&0&0&-126&0&-49518&2&0&-46&2\\
33&28530824630&-59178&1220&0&126&0&-6&12&0&0&0&2&-4&12&62&-1204&59430&0&0&-138&0\\
34&41286761478&70758&0&-72&-66&-10&-6&0&6&0&2&0&0&12&150&1332&-70890&0&2&54&0\\
35&59435554926&-84530&-1518&26&-154&6&2&-14&0&2&2&0&0&0&-66&0&84222&2&0&158&2\\
36&85137361430&100310&1670&0&70&-12&-2&-10&0&0&0&0&-2&-18&-170&-1666&-100170&0&0&-74&-2
\\
        \bottomrule
      \end{tabular}
 }
}
  \caption{Values $A_g(n)$ from twisted elliptic genera for lower levels $n$.}
  \label{tab:Fourier}
\end{table}

\begin{table}
  \newcolumntype{R}{>{$}r<{$}}
  \rowcolors{2}{gray!22}{}
  \rotatebox[]{90}{
    \resizebox{0.95\textheight}{!}{
      \centering
      \begin{tabular}{R||RRRRRRRRRRRRRRRRRRRRR}
        \toprule
        n & 1 & 23 &
        252&253&1771&3520&
        \displaystyle\frac{45}{45}
        &
        \displaystyle\frac{990}{990}
        &
        \displaystyle\frac{1035}{1035}
        &
        1035^\prime&
        \displaystyle\frac{231}{231}
        &
        \displaystyle\frac{770}{770}
        &483&265&
        2024&2277&3312&5313&5796&5544&10395
        \\
        \midrule \midrule
1&0&0&0&0&0&0&1&0&0&0&0&0&0&0&0&0&0&0&0&0&0\\
2&0&0&0&0&0&0&0&0&0&0&1&0&0&0&0&0&0&0&0&0&0\\
3&0&0&0&0&0&0&0&0&0&0&0&1&0&0&0&0&0&0&0&0&0\\
4&0&0&0&0&0&0&0&0&0&0&0&0&0&0&0&2&0&0&0&0&0\\
5&0&0&0&0&0&0&0&0&0&0&0&0&0&0&0&0&0&0&2&0&0\\
6&0&0&0&0&0&2&0&0&0&0&0&0&0&0&0&0&0&0&0&0&2\\
7&0&0&0&0&2&0&0&0&0&0&0&0&0&0&2&0&0&2&2&2&2\\
8&0&0&0&0&0&2&0&1&1&0&0&0&0&2&0&2&2&4&2&2&6\\
9&0&0&0&0&2&4&0&0&2&2&0&2&2&0&2&2&4&4&8&8&10\\
10&0&0&0&2&4&8&0&2&2&2&2&0&2&4&4&6&6&12&10&10&24\\
11&0&0&0&0&8&12&0&4&4&6&0&4&0&2&10&8&14&22&26&24&40\\
12&0&2&2&4&12&30&0&8&8&4&2&6&4&12&12&18&26&40&40&38&80\\
13&0&0&4&2&26&44&2&14&14&18&2&10&6&16&30&28&44&70&84&80&136\\
14&0&0&4&6&38&86&0&24&24&22&8&16&14&34&46&58&80&128&132&126&254\\
15&0&0&12&8&78&144&2&40&44&46&8&38&18&46&86&88&138&218&246&238&424\\
16&0&2&18&22&122&252&2&72&72&68&18&50&36&100&140&170&232&378&400&382&742\\
17&0&2&30&26&212&410&8&116&124&130&25&94&54&140&246&262&392&630&704&670&1222\\
18&0&6&50&58&342&704&6&194&202&192&50&148&100&256&388&454&654&1044&1120&1074&2058\\
19&0&4&80&72&582&1116&18&318&332&346&68&252&150&394&664&722&1062&1702&1880&1800&3320\\
20&0&14&128&138&904&1836&20&516&536&520&126&390&254&676&1036&1196&1716&2764&2980&2846&5408\\
21&2&20&214&200&1476&2902&40&814&860&872&182&652&396&1020&1684&1862&2742&4384&4828&4622&8572\\
22&2&32&328&346&2302&4616&55&1298&1348&1336&314&988&640&1686&2630&3000&4324&6950&7532&7204&13620\\
23&2&40&512&496&3638&7166&98&2020&2118&2144&460&1590&972&2546&4162&4624&6768&10856&11898&11376&21204\\
24&0&80&798&824&5584&11192&132&3140&3278&3236&744&2426&1544&4050&6376&7248&10500&16834&18294&17504&32976\\
25&8&108&1232&1208&8654&17084&234&4814&5038&5084&1106&3764&2336&6108&9892&11042&16112&25840&28288&27056&50524\\
26&6&174&1860&1904&13090&26148&322&7348&7670&7626&1742&5677&3602&9444&14968&16940&24566&39428&42894&41022&77176\\
27&12&252&2836&2802&19914&39436&514&11092&11618&11666&2560&8688&5394&14100&22744&25462&37148&59564&65114&62294&116494\\
28&16&398&4238&4310&29772&59330&742&16686&17418&17356&3922&12912&8160&21414&34026&38434&55764&89490&97456&93218&175146\\
29&26&560&6328&6286&44512&88280&1154&24840&25994&26078&5758&19380&12090&31636&50892&57068&83146&133356&145690&139342&260828\\
30&34&876&9368&9486&65776&131020&1642&36824&38480&38368&8642&28580&18008&47172&75158&84776&123176&197596&215318&205970&386724\\
31&58&1236&13802&13764&97060&192538&2500&54178&56660&56800&12582&42218&26384&69082&110920&124506&181274&290780&317502&303700&568798\\
32&76&1866&20166&20356&141714&282074&3564&79320&82884&82730&18576&61574&38738&101530&161978&182554&265284&425624&463950&443760&832834\\
33&122&2664&29396&29374&206524&410062&5286&115334&120644&120798&26830&89868&56226&147156&236010&265136&385974&619072&675796&646432&1211106\\
34&166&3900&42474&42810&298508&593800&7542&166990&174510&174330&39066&129694&81546&213644&341154&384250&558530&896052&977004&934530&1753356\\
35&248&5536&61184&61234&430134&854284&10988&240304&251292&251544&55956&187094&117138&306736&491602&552494&804038&1289768&1407604&1346380&2523178\\
36&334&8058&87622&88196&615626&1224424&15560&344314&359902&359564&80470&267604&168092&440318&703542&792158&1151786&1847690&2014952&1927370&3615350
        \\
        \bottomrule
      \end{tabular}
    }
  }
  \caption{Multiplicites $c_R(n)$ in decomposition of $A_g(n)$.}
  \label{tab:multiplicity}
\end{table}

\clearpage


\newpage


\end{document}